\documentclass{ws-ijmpcs}

\usepackage[active]{srcltx}

\usepackage{amsmath,amssymb,amsfonts,amsbsy}
\usepackage{graphicx}
\usepackage{wrapfig}
\usepackage{epsfig}
\newcommand{\be}{\begin{equation}}
\newcommand{\ee}{\end{equation}}
\newcommand{\ba}{\begin{eqnarray}}
\newcommand{\ea}{\end{eqnarray}}

\newcommand{\di}{ {\rm d} }
\usepackage{color}
\definecolor{darkgreen}{rgb}{0,0.65,0}

\usepackage{amsfonts}
\usepackage{amssymb,latexsym}
\usepackage{amsmath,amsbsy}

\def\trimmarks{}

\begin{document}
\markboth{A.V. Efremov, I.F. Ginzburg and A.V. Radyushkin}
{Regge trajectories in QCD}

\title{REGGE TRAJECTORIES IN QCD
}

%
\author{\underline{A.V. EFREMOV}}

  \address{Joint Institute for Nuclear Research, Dubna,
141980 Russia} 

\author{I.F.  GINZBURG}
  \address{Sobolev Institute of Mathematics SB RAS, pr. ac. Koptyug, 4 and Novosibirsk State University,
Novosibirsk 630090, Russia} 

\author{A.V. RADYUSHKIN}
  \address{Old Dominion University,
Norfolk, VA 23529 and Thomas Jefferson National Accelerator
Facility, Newport News, VA 23606, USA}

\maketitle

\begin{abstract}
We discuss some problems concerning the application of
perturbative QCD to high energy soft  processes. We show
that summing the contributions of the lowest twist operators for
non-singlet $t$-channel leads to a Regge-like amplitude. Singlet
case is also discussed.


\end{abstract}

\ccode{PACS numbers: 12.38.-t, 12.38.Cy, 12.39.St, 11.55.Jy}


\vspace{0.5cm}

The famous asymptotic freedom of  quantum chromodynamics (QCD)
[\refcite{ref1}] enables one to use perturbation theory (PT) for large
momenta. However, any physical process also involves small
momentum scales $p^2$ (e.g., quark and hadron masses). As a rule,
this results in logarithmic contributions $\ln(Q^2/p^2)$ singular
for $p^2=0$ (mass singularities) [\refcite{ref2}]. In such a situation
$p^2$ cannot be neglected. Within PT it is possible to show that
for inclusive [\refcite{ref2}]--[\refcite{ref6}] and some exclusive hard
processes (see, e.g., [\refcite{ref6,ref7}]) the $Q^2$-dependence of
the corresponding amplitude $T(Q^2,p^2)$ can be factorized from
the $p^2$-dependence
\be
T(Q^2,p^2)= Q^N\left\{E(Q^2/\mu^2,\alpha_s(\mu))\otimes
f(\mu^2,p^2) +R(Q,p)\right\} \ , \label{eq1.1}
\ee
where $R$ is sum of power suppressed contributions. Note, that
$E\otimes f$ does not depend on a particular choice of  $\mu$, the
boundary between large and small momenta.

Our aim here is to recall\footnote{This talk was motivated by
papers [\refcite{Efremov:2009dx}],  [\refcite{Gin}].} that a similar
approach is also possible  for soft binary processes ${\bf 12 \to
1'2'}$ in the region $S\equiv (s-u)/2\gg|t|,\,m_{\rm hadr}$ and
attract  attention to  some dangerous points in widespread
constructions. Just like in our derivation of factorization
formula (\ref{eq1.1}) in
Refs.~[\refcite{ref5}]--[\refcite{Efremov:2009dx}], we start with the
$\alpha$-representation [\refcite{ref17}] for the amplitude
\begin{align}
\hspace{-3mm} F(S,t)\sim &\int\limits_0^\infty
\prod\limits_\sigma \di\alpha_\sigma\,D^{-2}(\alpha)\,
G(S,t,\alpha)  \exp\left[ iS A(\alpha)+i t I(\alpha) -
i\sum_\sigma \alpha_\sigma m_\sigma^2 \right]  \,  ,
\label{eq3.2}
\end{align}
which has many advantages for analysis  of the large $S$ behaviour
of $F$.

In particular, according to (\ref{eq3.2}), integration over a
region where $A(\alpha) > \rho$  gives for $S\to \infty$ an
exponentially damped contribution ${\cal O}[\exp (-S \rho)]$.
Hence, contributions having a power ${\cal O}(S^{-N})$  behaviour
for $S \to \infty $ are due to integration over regions where
$A(\alpha)$ vanishes. Three main possibilities to get
$A(\alpha) = 0$  are: %
 1) short-distance (SD, or small-$\alpha$)  regime, when
$\alpha_{\sigma_1}= \alpha_{\sigma_2} = \ldots =
\alpha_{\sigma_n} = 0$ for some lines $\sigma_1, \sigma_2, \ldots, \sigma_n$;
 2) infrared  regime, when $\alpha_{\sigma_1}=
\alpha_{\sigma_2} = \ldots = \alpha_{\sigma_n} = \infty$ for a
set of lines $\{ \sigma_1, \sigma_2, \ldots , \sigma_n \}$;
 3) pinch regime, when $A(\alpha)=0$ for nonzero finite $\alpha$'s
because  $A(\alpha)$ is a difference of two positive terms.

In  a wide class of processes, the pinch regime does not work
(see Refs.~[\refcite{ref5}], [\refcite{ref45}]). Since for $A(\alpha) = 0$
the amplitude $F$ lacks its $S$-dependence, one must find the
subgraphs $V,L$ with the property that, when lines of  $V$
subgraphs are contracted into point ($\alpha_\sigma= 0$) and/or
lines of $L$ subgraphs are removed ($\alpha_\sigma= \infty$), the
resulting diagram does not depend on $S$. The power $N$ of  such
a ${\cal O}(S^{-N})$ contribution may be easily estimated by
``twist counting rules'': %
 $%
 t_V^{\rm SD}  \lesssim S^{4 -\sum t_i} \ ;\
 t_L^{\rm IR}  \lesssim S^{ -\sum t_j} \ ;\
 t_{V;L}^{\rm SD;IR}\lesssim S^{4 -\sum t_i-\sum t_j}\ ,
 $ %
where $t_i$  ($t_j$) is twist of the $i$-th ($j$-th) external
line of the subgraph $V$ ($L$). Since $t_{i,j} =1$  for
$\psi,\bar \psi$-fields and the field strength $G_{\mu \nu}$,
whereas  $t_{i,j} =0$  for the vector potential  $A_\mu$, it is
necessary in QCD (in covariant gauges) to sum up over external
gluon lines of the subgraphs $V, L$.

\begin{figure}
\centering
\includegraphics[width=.85\textwidth]
{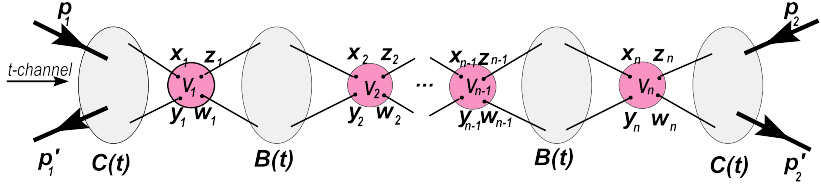}
\caption
{ Illustration of Equation (\ref{eq5.1}).}
\label{fig7}
\end{figure}

In a Yukawa-type theories it has been shown 40 years ago
[\refcite{ref45}] that, for amplitudes with positive signature, the
logarithmic terms $(\log S)^N$  appear only from the SD
integration, and their summation gives the following
representation (see \mbox{Fig. 1})
\be
{{f^+(j,t)=C^{T}(j,t)[1-B(j,t) v(j)]^{-1}
v(j)C(j,t)}}
\label{eq5.1}
\ee
for the Mellin transform of the scattering amplitude
$F^{\pm}(S,t)$:
\be
F^{\pm}(S,t)=\frac{1}{2i}\int\limits_{-i\infty}^{i\infty}\!\!\!\di
j \,\frac{|S|^j(e^{i\pi j}\pm 1)}{\Gamma(j+1)\sin(\pi
j)}f^\pm(j,t)  \  , \label{eq5.2}
\ee
where $\pm$ stands for signature, $C,\,v\mbox{ and } B$ are 
matrices $2\times 2$ or $3\times 3$. According to the
representation (\ref{eq5.1}), the Mellin transform $f^\pm(j,t)$
possesses moving ($t$-dependent) Regge poles due to zeros of
${\rm Det} [1-B(j,t)v(j)]$. It has also fixed ($t$-independent)
singularities in the complex $j$-plane accumulated in the
function $v(j)$. The type of fixed singularities depends on the
ultraviolet behavior  of the effective coupling constant. {{ Note
that $v(j)$ has the form of LLA result, corrected by next
approximations, while Regge pole structure cannot be seen from
NLLA, NNLLA, etc. }}

The Mellin transform of Eq. (\ref{eq5.2}) has the following
structure in the $\alpha$-representation
 \be
f^\pm(j,t)\propto \int\prod\limits_\sigma\di\alpha_\sigma
D^{-2}(\alpha)g(j,t,\alpha)|A(\alpha)|^j [\theta(A)\pm\theta(-A)]
\exp[i J(\alpha,t,m^2)]\,,
 \label{eq5.3}
 \ee
where $g(j,t,\alpha)$ is a polynomial in $j$ (it corresponds to
the function $G$ in Eq. (\ref{eq3.2})) and $A$ is the coefficient
in front of the large variable $S$. The asymptotic behavior of
$F(S,t)$ for large $S$ is determined by the rightmost
singularities of its Mellin transform $f(j,t)$. These are the
poles $1/(j-N)$ generated by integrations corresponding to the
regimes  discussed earlier. By the twist counting rules given
above, the leading  poles (at $j=0$) are due to the SD  subgraphs
$V_i$ with 4 external lines, since the IR-regime   in a Yukawa
theory gives only non-leading poles at $j=-1,-2,\dots$.
Furthermore it was proven [\refcite{GinSas}] that for even $j$  the
pinch regime contributes only to the negative signature amplitude
$F^-(S,t)$, while for the odd $j$ -- only to positive signature
amplitude $F^+(S,t)$. That is the reason why for $F^+(S,t)$ it is
sufficient to consider only the  SD  poles at $j=0$ (Eq.
(\ref{eq5.1})).

\begin{figure}
\centering
\includegraphics[width=.65\textwidth]{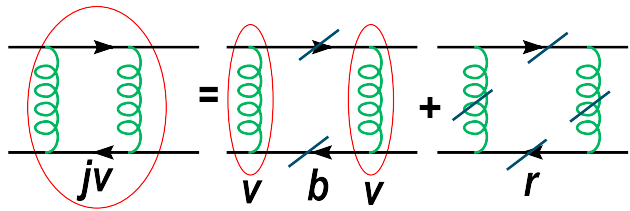}
\vskip3mm
\caption
{
The leading terms of a
particular  box diagram. } \label{fig8}
\end{figure}

In general, the SD-subgraphs $V_i$ may contain smaller
$V$-subgraphs with 4 external lines, and the total singularity
due to the SD-regime of $V_i$ may be a multiple pole $j^{-N_i}$.
Treating  a particular diagram as a ladder composed by 2-particle
irreducible blocks $k_j$, we see that  {\it the maximal} value of
$N_i$ is determined by the number of $k_j$'s inside $V_i$ (and
also by the number of the UV-divergent subgraphs inside $V_i$).
In Fig. 2, the pole parts are circled by the thin (red) line and
the regular ones are marked by the blue lines. It immediately
gives equation $ \mu \frac{dv}{d\mu}=jv=r+bv^2\, , $ where $b$
and $r$ {{are}} some functions regular at $j=0$. Using this
equation, one can sum up all leading poles at $j=0$ due to the
SD-regime of all possible $V$-subgraphs (for summation of all
poles see  Ref. [\refcite{Efremov:2009dx}]), i.e. to sum all leading
$\log^N(S/p^2)$ contributions. The solution has square root
branch points in the complex $j$-plane [\refcite{ref45}] (see also
Ref.~[\refcite{ref46}]). However, $v(j)$ has also  poles due to
divergent subgraphs. These poles (i.e.
$\log^N(S/\mu_R^2)$-contributions) are summed by the
renormalization group equation
\be
(\mu_R\partial/\partial\mu_R+\beta(g)\partial/\partial g-
4\gamma_\psi)v(j,\mu_R,g,\mu)=0 .
\ee
Generally $\mu_R$ and $\mu$ are quite different scales but if
we take $\mu=\mu_R$, we obtain the combined equation %
$\beta(g){\partial v}/{\partial g}=(j-2\gamma-4\gamma_\psi)v-
b v^2-r\, . $ %
In the lowest order of PT $b=1,\, r\sim\gamma \sim\gamma_\psi\sim
g^2$, $\beta(g)\sim g^3$, and the solution of this equation has
condensing poles at $j=0$ [\refcite{ref45,ref46,GS}].

Summarizing, in addition to a  fixed singularity near $j=0$, the
amplitude $F(S, t)$  has  a Regge-type behaviour $F(S,t)\sim
C^2(t)S^{\alpha(t)}$ for large $S$. To find the function
$\alpha(t)$ explicitly, one must solve the equation \mbox{$ {\rm Det}[1
-B(j,t,\mu_R,g,\mu,m)v(j,\mu_R,g,\mu)]=0\,. $} In fact
[\refcite{Efremov:2009dx}], $\alpha(t)$ does not depend on $\mu$ and
$\mu_R$: %
$ \alpha(t)=
\phi(m_q^2/t,t/\mu^2,\bar{g}(\mu^2))=\phi(m_q^2/t,1,\bar{g}(t))
$. %
Hence, one may try to calculate the Regge trajectories in the
region where $g(t)$ is small, e.g.  in QCD for
sufficiently large $t$.

\begin{figure}
\begin{center}
\vspace{-3cm}
\hspace{-1.5cm}
 \includegraphics[width=.58\textwidth]{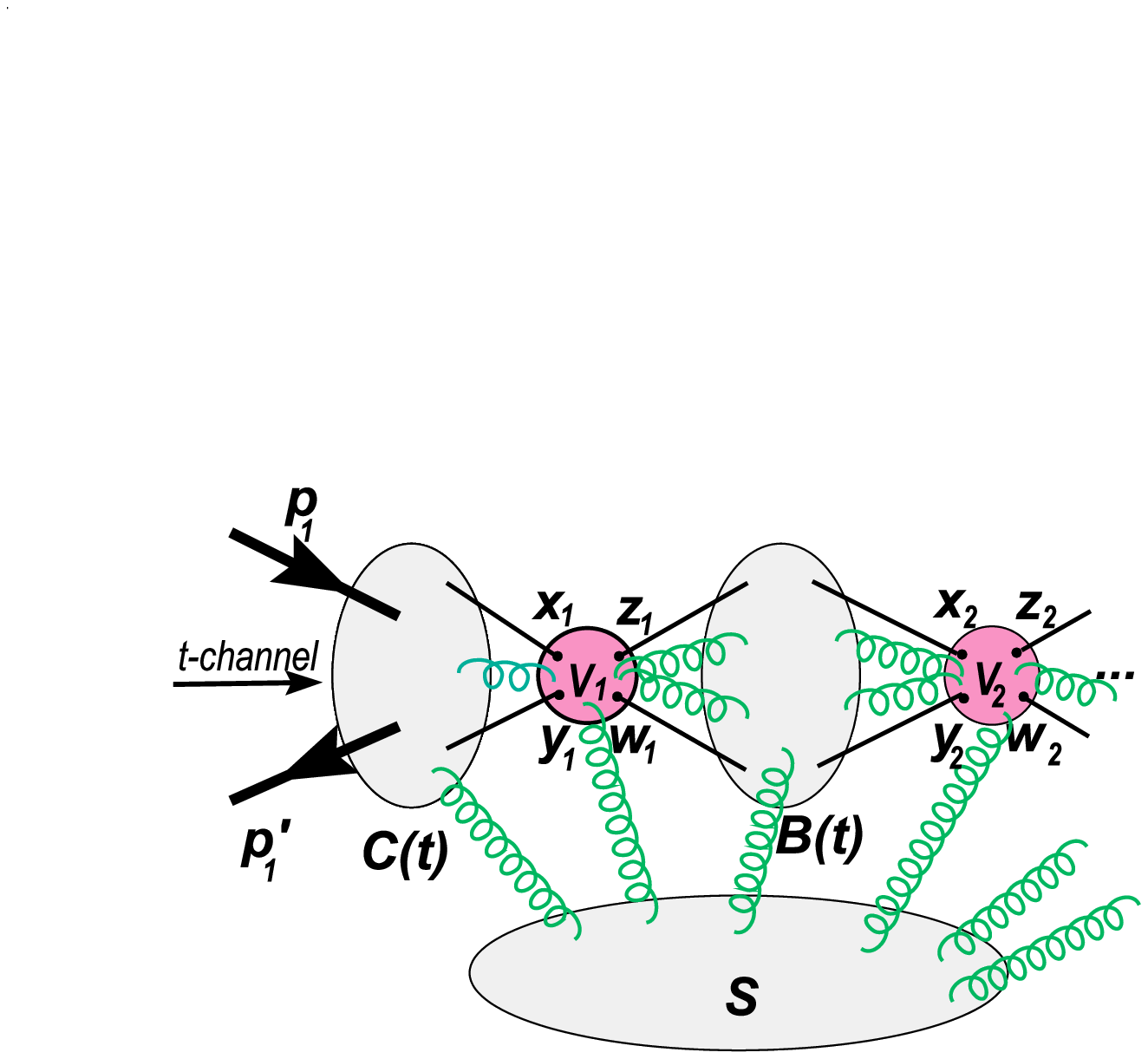}
\end{center}
\caption
{The same as Fig.
1\ref{fig7}, but for the case of QCD.} \label{fig9}
\end{figure}

In QCD, for non-singlet $t$-channel one faces  complications
discussed earlier. First,  SD-subgraphs $V_i$ may have an
arbitrary number of external gluon lines. Still,  if the
\mbox{$t$-channel} is color singlet,   the only change is (see
e.g. Ref. [\refcite{ref5}]) a  path-ordered exponential between the
$x$ and $y$ points for all bilocal operators
$\bar\psi(x)\Gamma\psi(y)$ entering into $B$- and $C$-functions.
For local operators,  this corresponds to the change
$\partial_\mu\to D_\mu=\partial_\mu-i g\hat A_\mu$. The second
complication is due to the IR-regime (soft exchanges, see Fig.
3). However, if the $t$-channel is color singlet, then the sum of
all soft exchanges gives  only power corrections in each order of
perturbation theory. Thus, all terms responsible for the leading
power contribution have the structure of Fig.~\ref{fig7} and as a
result, we get Eq.~(\ref{eq5.1}).

Note, that the function $B(t)$ describes the long-distance
dynamics, and one must take into account nonperturbative effects,
e.g., using  QCD Sum Rules approach [\refcite{ref26}] that assumes
non-zero  vacuum expectation values of some products of field
operators  (vacuum condensates). Then  only equations for $B$ and
$C$ are changed. However, one cannot tell  what kind of
contributions dominates: fixed singularity in $j$-plane (LLA
type) or a Regge pole. As it is known the Regge poles are
dominant for nonsinglet $t$-channel. In this sense,  improvements
of LLA, like including next-to-leading logs,
 could be misleading!

We have discussed above {\it the flavor nonsinglet, positive
signature amplitude $F^+_{NS}$ only}. For $F^-_{NS}$,  the pinch
regime also gives leading $j$-poles for non-planar diagrams.
However, in QCD the non-planar diagrams have an additional color
factor $(1/N_c)^2=(1/3)^2$. This suggests that the pinch
contributions in QCD may be suppressed.

For flavor singlet amplitudes $F_S$ in QCD, the poles generated
by the pinch regime are at $j=1$ rather than at $j =0$ due to the
2-gluon intermediate states. As a result, the  asymptotic
behavior of Pomeron amplitude $F^+_S$ is determined by a
complicated  mixture of SD and pinch singularities. The LL
results were obtained starting with Ref. [\refcite{BFKL}]. Usefulness
of recent NLL results is  not very clear for us due to reasons
discussed above (namely, what is eventually the leading
contribution).

On the other hand, the asymptotic behavior  of the odderon
amplitude $F_S^-$ has a much simpler structure: it is determined
by small distance singularities at $j=1$ only. The corresponding
direct analysis, starting from diagrams, was not done till now.
Nevertheless,  the representation {\it odderon = colored Pomeron
+ gluon} looks misleading for us.

In conclusion, a lot of important and interesting things are yet
not solved in QCD of soft high energy processes, which  asks for new
young forces and new ideas.

\section*{ Acknowledgement} 
{Tne paper is supported by RFBR
grants 11-02-00242 and 12-02-00613;
the work of A.R. 
is supported by Jefferson Science Associates, LLC under  U.S. DOE Contract No. DE-AC05-06OR23177.  
}


\end{document}